\documentclass[aps,a4paper,10pt,twocolumn,footinbib]{revtex4} 
\usepackage{datetime}
\usepackage{amsmath}
\usepackage{amsfonts}
\usepackage{amsfonts}
\usepackage{mathrsfs}
\usepackage[mathscr]{euscript}
\usepackage[dvips]{graphicx}
\usepackage{fancyhdr}
\usepackage{colordvi}
\usepackage[hypertex=true]{hyperref}
\usepackage{epsfig}
\usepackage{color}

\def\overstrike#1#2{{\setbox0\hbox{$#2$}\hbox to \wd0{\hss
    $#1$\hss}\kern-\wd0\box0}}

\newdateformat{yymmdddate}{\THEYEAR/\twodigit{\THEMONTH}/\twodigit{\THEDAY}}

\begin{document}
\title{Comment on: 
  Reply to comment on 
    `Perfect imaging without negative refraction'}
\author{Paul Kinsler}
\email{Dr.Paul.Kinsler@physics.org}
\affiliation{
  Blackett Laboratory, Imperial College,
  Prince Consort Road,
  London SW7 2AZ, 
  United Kingdom.
}
\author{Alberto Favaro}
\affiliation{
  Blackett Laboratory, Imperial College,
  Prince Consort Road,
  London SW7 2AZ, 
  United Kingdom.
}

\lhead{\includegraphics[height=5mm,angle=0]{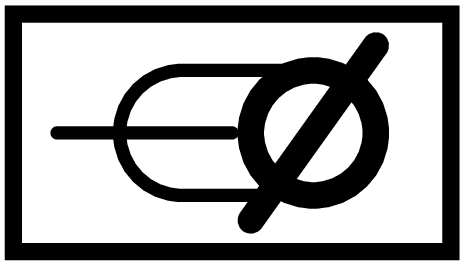}}
\chead{On `Perfect imaging without NR'}
\rhead{
\href{mailto:Dr.Paul.Kinsler@physics.org}{Dr.Paul.Kinsler@physics.org}
}
\rfoot{{\large \emph{Published in New J. Phys. \textbf{13}, 028001 (2011)}\\
doi:10.1088/1367-2630/13/2/028001}}

\begin{abstract}

Whether or not perfect imaging is obtained 
 in the mirrored version of Maxwell's fisheye lens
 is debated
 in 
 the comment/reply sequence 
 \cite{Blaikie-2010njp-fisheyeC,Leonhardt-2010njp-fisheyeR}
 discussing Leonhardt's original paper \cite{Leonhardt-2009njp}.
Here we show that
 causal solutions \emph{can} be obtained without the need for 
 an ``active localized drain'',
 contrary to the claims in \cite{Leonhardt-2010njp-fisheyeR}.

\end{abstract}

\date{\today}
\maketitle
\thispagestyle{fancy}

\noindent{\emph{Published in New J. Phys. \textbf{13}, 028001 (2011).\\
\footnotesize{\url{http://iopscience.iop.org/1367-2630/13/2/028001}}}\\
\scriptsize{\textbf{Statement required by the publisher of the NJP:}
 This is an author-created, un-copyedited version of an article
  accepted for publication in the New Journal of Physics. 
 IOP Publishing Ltd is not responsible for any errors or omissions
  in this version of the manuscript or any version derived from it. 
 The definitive publisher-authenticated version is available online
 at doi:10.1088/1367-2630/13/2/028001.}
}

%

%

~\\
R.J. Blaikie (RJB) notes in \cite{Blaikie-2010njp-fisheyeC} that 
 that Ulf Leonhardt's (UL) setup in \cite{Leonhardt-2009njp}
 incorporates an ``active localized drain'' at the 
 image point.
It is this drain, 
 modeled as a phase-delayed mirror image of the source, 
 which provides the sub-wavelength detail of the source's image.
RJB showed that a steady-state numerical simulation of the fields
 without the drain did not show
 the perfect super-resolution image of the source.
In response, 
 UL noted \cite{Leonhardt-2010njp-fisheyeR} 
 that steady-state solutions neglect causality
 and that inclusion of a sink solves the problem of energy buildup.

\begin{figure}[h]
\includegraphics[width=0.4\columnwidth,angle=0]{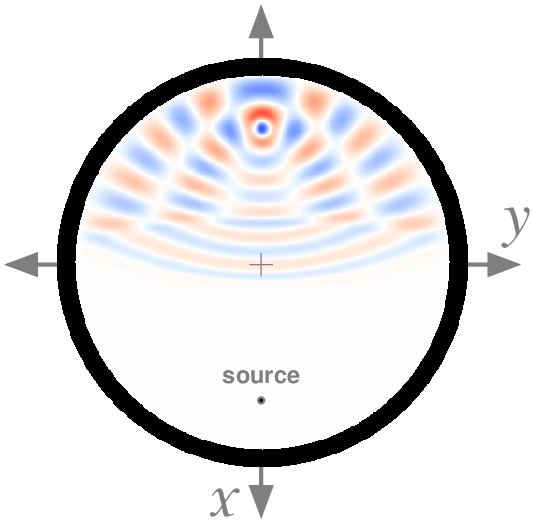}
\caption{
Snapshot of the electric field $E_z$ 
 from a simulation of the cylindrical mirrored fisheye, 
 as the light first reaches the image point.
Parameters are taken from RJB but with zero losses:
 $n(r)=2/(1+(r/r_0)^2)$ for $r_0=10\mu$m,
 $f=100$THz (i.e. $\lambda_0=3\mu$m).
}
\label{fig-xy}
\end{figure}

\begin{figure}
\includegraphics[width=0.8\columnwidth,angle=0]{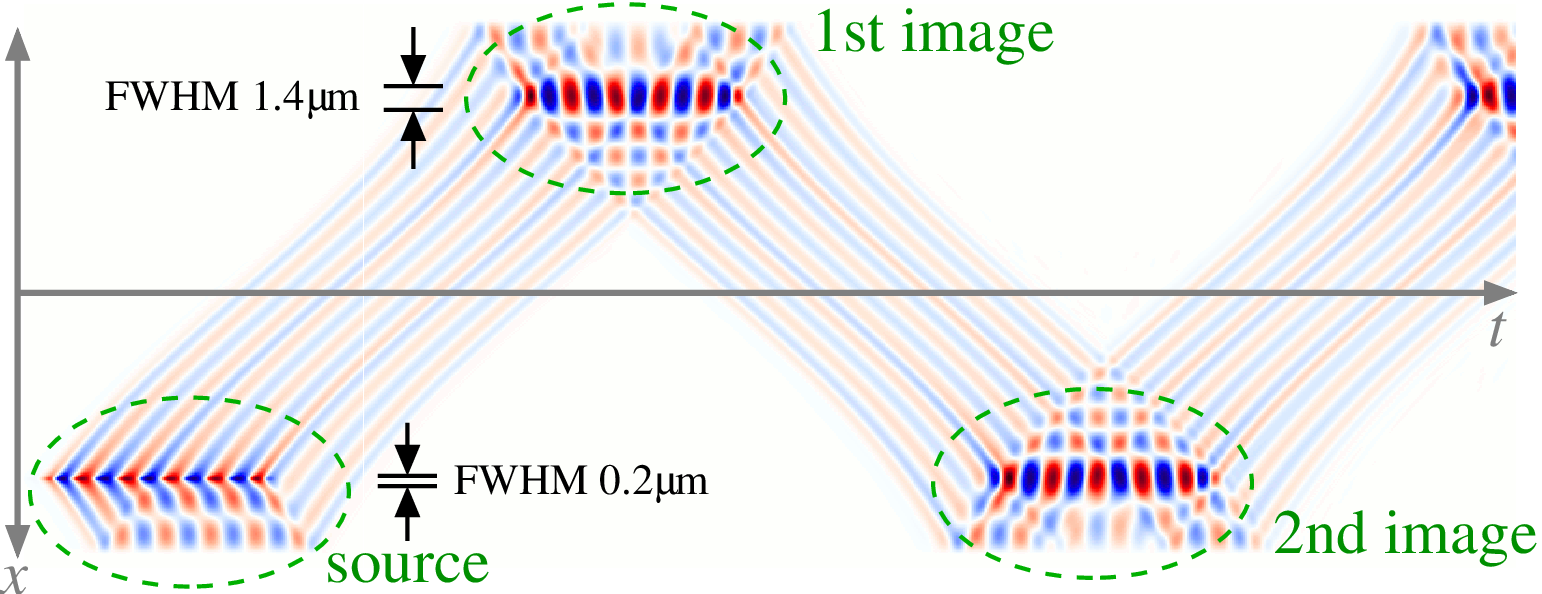}
\caption{
Time history of the simulation in fig. \ref{fig-xy}
 showing the field $E_z$ along the $x$-axis.
The intensity FWHM of the source and image(s) are given, 
 and can be compared to either $\lambda_0$ or the 
 local wavelength ($\lambda_0/n\simeq 2.3\mu$m) at the image point.
Further bounces gradually degrade the image quality, 
 and more so for briefer source durations.}
\label{fig-yslice}
\end{figure}

We now address both of UL's concerns about RJB's simulations
 using a \emph{time-domain} numerical solution
 with a source only active for a finite time.
In figs.~\ref{fig-xy} \& \ref{fig-yslice}
 the results from such a simulation, 
 calculated using the open source MEEP \cite{Oskooi-RIBJJ-2010cpc}
 implementation of FDTD \cite{Taflove-FDTD},
 are shown. 
In FDTD, 
 the real valued electric and magnetic fields
 are defined over space, 
 and an algorithm is applied which propagates these fields,
 step by step, 
 forwards in time.
It is thus explicitly causal, 
 and entirely independent of any decomposition
 into plane waves or modes.

The point-like source used is independent of all other 
 properties of the simulation\footnote{If
  there were e.g. 
  a spatially separated source/ active drain pair, 
  we would have to ensure that any desired correlations 
  between them remained consistent with causal signalling.}.
We follow UL and RJB
 and use a frequency independent refractive index profile, 
 so there is no temporal dispersion.
Our simulation therefore correctly tests for the geometric response
 and achievable spatial resolution, 
 irrespective of the fact that transients are used
 instead of steady-states.
Fig. \ref{fig-yslice} shows
 that the image is \emph{not} as sharp as the source, 
 and matches the steady-state results
 of RJB \cite{Blaikie-2010njp-fisheyeC}.

We do not include an active drain in our simulations because
 we believe it unlikely to form a part of any actual device; 
 e.g. if this mirrored fisheye replaced
 the elliptical cavity used in lamp-pumped lasers.
Active drains often appear in the literature 
 when systems with sources are designed using 
 folded-space or mirror-imaged transformations:
 e.g. the transformation optics slab-lens
 \cite{Leonhardt-P-2006njp}.
However some authors
 insist that such active drains are unphysical
 and/or mathematically ambiguous 
 \cite{Maystre-E-2004josaa,Merlin-2004apl,Milton-NMP-2005rspa,
       Bergamin-F-2010arxiv}.
RJB was able to replace the drain
 in his \emph{steady-state} simulation
 with a carefully-phased source \cite{Blaikie-2010njp-fisheyeC}, 
 but this will fail in general for both time domain simulations
 and physical devices \cite{Kinsler-2010pra-lfiadc}.
Whatever the method, 
 we consider achieving super-resolution by such means
 to be of little utility, 
 since it requires a customised element
 to enhance and `image' the field at each precisely tuned pixel.

In summary,
 despite the claims in \cite{Leonhardt-2010njp-fisheyeR},
 causality does not require the presence of an active drain, 
 irrespective of whether or not an active drain
 might be otherwise useful.

%

\begin{widetext}

\section*{Appendix: Sample MEEP control file}

\begin{verbatim}

; Maxwell fisheye in 2d
; 
; Paul Kinsler / Imperial College London,   May 2010
; 
;
;
; Treat sizes as if in microns (hence radius 10 ==> radius 10um)

; ======================================================================
; Fisheye sizes and parameters
; 
(define-param n  2)    ; base index of fisheye
(define-param w  1)    ; width of waveguide
(define-param r 10)    ; inner radius of ring

; ======================================================================
; Computational resolutions etc
;
(set-param! resolution 20)  ; cells per unit size
                            ; NB: 10 is ok for a couple of bounces
                            ;     but increase it for longer sims
(define-param pad 2)        ; padding between waveguide and edge of PML
(define-param dpml 2)       ; thickness of PML

; Here we make sxy non-integer so that the simulation will have an 
; odd-number x and y cell count; this is because I want an exact on-axis
; readout of fields etc without having to average array elements. 
; 
(define sxy (+ (/ 1 resolution) (* 2 (+ r w pad dpml))))
(set! geometry-lattice (make lattice (size sxy sxy no-size)))


; ======================================================================
; Refractive index profile functions 
;  to calculate the local refractive index as it varies with position.
;

; DEFINE frr = sqrt(x^2 + y^2)
;
(define (frr p)
        (sqrt (+ (* (vector3-y p) (vector3-y p)) 
                 (* (vector3-x p) (vector3-x p))
)       )     )

; DEFINE fdivisor = 1 + rr^2/r^2
;
(define (fdivisor rr)
   (+   1   ( / (* rr rr) (* r r) )
)  )

; DEFINE refindex = n / fdivisor = n / ( + rr^2/r^2)
;
(define (refindex rr)
   (/ n (fdivisor rr))
)


; DEFINE eps = refindex^2 = the relative permittivity
;
(define (eps rr)
   (* (refindex rr) (refindex rr))
)

; DEFINE fmedium (x,y)
; 
(define (fmedium p)
        (make medium
                (epsilon (eps (frr p)))
)       )




; ======================================================================
; 
; Create the Fisheye structure
;

; Create a metal ring to form the mirror using two overlapping cylinders 
;  - later objects take precedence over earlier objects, so we put the 
; outer cylinder first, and the inner (air) cylinder second.
; 
; The first three blocks are just to add some guides for the 
;   eye locating the source and image points --
;   they have no effect inside the mirrored fisheye
;
(set! geometry 
  (list 
        (make cylinder (center 0 0) (height infinity)  
                       (radius (+ r w)) 
                       (material perfect-electric-conductor)
                       (material perfect-magnetic-conductor))
        (make cylinder (center 0 0) (height infinity) 
                       (radius r) 
                       (material (make material-function 
                                       (material-func fmedium) )))
;
;   Some objects we might like to try putting at the image point:
;
;        (make cylinder (center -7.20 0) (height infinity)  
;                       (radius 0.02) 
;                       (material metal)
;                       ;(material perfect-magnetic-conductor)
;                       ;(material (make dielectric (epsilon 1.000001)
;                       ;             (E-polarizations 
;                       ;             (make polarizability
;                       ;             (omega 0.01) (gamma 1) (sigma 1.0e5)))))
;        )
) )


; ======================================================================
; Set up the PML at the simulation boundaries
;

(set! pml-layers (list (make pml (thickness dpml))))

; ======================================================================
; 
; SOURCES: Put a single point source on the y-axis at x=7.20um
; 
; 

(define-param  ffq       (/   1  3))     ; scaled inverse wavelength of 3um
(define-param  fdelta    (* ffq 20.25))  ; a time delay that might be useful
(define-param  fduration (/  6  ffq))    ; a on/off time scale for switching

(set! sources 
  (list
;   Finite-time source at x=7.20um with gradual switch on/off
;      to avoid generating high frequency components that will
;      clutter the output (but would propagate just the same 
;      in this dispersionless system).
;
    (make source
      (src (make continuous-src (frequency ffq) (width (* 2 ffq))
                                (end-time fduration)))
      (component Ez) (center  7.20 0.00)
    )
;
;   Trial (pseudo-drain) source
;
;    (make source
;      (src (make continuous-src (frequency ffq)  (width (* 2 ffq)) 
;                                (amplitude -1)
;                                (start-time fdelta) 
;                                (end-time (+ fduration fdelta))))
;      (component Ez) (center -7.20 0.00)
;    )
  )
)


; ======================================================================
; 
; Run the simulation
; 
; 

(run-until 132
           (at-beginning output-epsilon)
           (to-appended "I" 
             (at-every 0.125 (synchronized-magnetic output-tot-pwr))
           )
           (to-appended "ez" 
             (at-every 0.125 output-efield-z))
           )


\end{verbatim}
\end{widetext}

\end{document}